**Original Manuscript**

**Title**

Countermeasure against Side-Channel Attack in Shared Memory of TrustZone

Na-Young Ahn[1], and Dong Hoon Lee[2]

*1. Na-Young Ahn is with Graduate School of Information Security, Korea Universiy, Seoul, Republic of KOREA,  E-mail: humble@korea.ac.kr.*
 *2. Dong Hoon Lee is a Professor of CIST, Korea University, Seoul, Republic of KOREA, E-mail: donghlee@korea.ac.kr.*

**Abstract**

In this paper we introduced countermeasures against side-channel attacks in the shared memory of TrustZone.  We proposed zero-contention cache memory or policy between REE and TEE to prevent from TruSpy attacks in TrustZone. And we suggested that delay time of data path of REE is equal or similar to that of data path of TEE to prevent timing side-channel attacks. Also, we proposed security information flow control based on the Clark-Wilson model, and built the information flow control mechanism using Authentication Tokenization Program (ATP). Accordingly we can expect the improved integrity of the information content between REE and TEE on mobile devices.

**Introduction**

TrustZone is hardware-based security built into system-on-chips (SoCs) by semiconductor chip designers. Main theme of TrustZone is the concept of secure and non-secure worlds that are hardware separated, with non-secure software blocked from accessing secure resources directly [1], [2]. Within the processor, software either resides in the secure world (or trusted execution environment TEE) or the non-secure world (rich execution environment REE); a switch between these two worlds is accomplished via software referred to as the secure monitor or by the core logic. This concept of TEE and REE extends beyond the processor to encompass memory, software, bus transactions, interrupts and peripherals within SoC [3].

Figure 1 illustrates TrustZone's software architecture in a mobile device. REE's mobile OS accesses TEE via a TrustZone library and a hardware driver. In the TEE, trusted applications execute on top of a minimal runtime environment, called the trusted OS, which provides a TEE internal application program interface (API) that trusted applications can use for communication with REE applications to access cryptographic operations and secure storage functionality. The trusted OS can enforce access control on trusted applications that attempt to access the secure memory [4], [5], [6]. TrustZone architecture does not define how REE applications access TEE [4].

**Shared Memory in TrustZone**

Typically, TEE includes three types of hardware models: 1) separated hardware REE and TEE 2) integrated REE and TEE 3) REE and TEE having shared hardware [7]. TrustZone is configured to include the shared hardware, referring to figure 2. Access of the shared hardware is determined according to the encrypted bit NS-bit.

Particularly, REE and TEE share a main memory, for example, Dynamic Random Access Memory (DRAM). That is, DRAM is the memory shared by REE and TEE. NS-bit indicates whether to access one of the nonsecure area of DRAM and the secure area of DRAM [8]. The secure area is a space to store data encrypted by crypto algorithms.

**Side-channel Attack in Shared Memory**

Recently, Ning Zhang, Kun Sun, Deborah Shands, Wenjing Lou, and Y. Thomas Hou addressed the side-channel attack against the shared memory of TrustZone [9]. This side-channel attack is referred as TruSpy, which is the first study of timing based cache side-channel information leakage of TrustZone. TruSpy attack exploits the cache contention between REE and TEE as a cache timing side channel to extract sensitive information from the secure world, referring to figure 3. There are two attack requirements for the TruSpy attack. Firstly, the attacking process can fill in cache lines at individual cache sets that will cause cache contention between REE and TEE. Secondly, the attacker can detect the state change in the cache lines [9].

The TruSpy attack scenario consists of five steps below [9], [10], referring to Figure 4:

The first step is to identify the cache memory to use for cache priming. The key is to find the cache memory that will be filled in cache line that is also used by the victim process in TEE. This step is often accomplished by working out the mapping from virtual address to cache lines [9].

The second step is to fill all cache lines. The spy process fills the cache with its own memory so that each cache line that can be used by the victim is filled with memory contents from the address space of the attacker [9].

The third step is to trigger the execution of the victim process in TEE. When the victim process is running, cache lines that were previously occupied by the attackers are evicted to the cache memory, such as a DRAM. As a result, the cache configuration from the attacker's perspective has changed because of the execution of the victim process. Since this step is non-interruptible due to the protection of TrustZone, it is more challenging for this attack to succeed without fine grained information on the victim process cache access [9].

The fourth step is to measure the state change in cache configuration after the victim finishes its execution in TEE. For each cache line that was previously primed in the second step, the time to execute memory load instruction is measured. If the time it takes to load the cache memory into register is short, then cache lines of which the cache memory is mapped to was not evicted by the victim process. Once the results are recorded for all the cache memory locations that were primed, the attack goes back to the second step and continues to collect more side-channel information [10].

The fifth step is to analyse the collected channel information to recover secret information such as cryptographic keys within the secure domain.

**Proposed Countermeasures against TruSpy**

In this paper, the proposed countermeasure against the TruSpy side-channel attack addresses largely two points. One point is to remove or mitigate cache contention about the shared memory. Another point is to adjust timing of data paths for non-detecting cache transition.

**Zero Contention in Cache Lines**

In the above TruSpy modelling, this attack begins due to contention regarding shared cache lines between REE and TEE. Accordingly, if we remove the contention between REE and TEE, TruSpy side-channel attack can be blocked.

**Hardware Zero Contention Scheme**

To achieve zero contention between REE and TEE, we suggest a separate cache memory hardwaredly. In an embodiment, the cache memory is divided from fixed

REE-only cache lines and fixed TEE-only cache lines to remove or reduce the contention between REE and TEE with respect to the shared cache lines. For example, the fixed REE-only cache lines and the fixed TEE-only cache lines are hardware-implemented in the cache memory to achieve the zero contention, referring to figure 5.

**Software Zero Contention Scheme**

In another embodiment, an allocation of REE addresses and TEE addresses is fixed according to the cache policy. REE addresses and TEE address are separated from each other, referring to figure 6. REE addresses correspond first cache lines in the cache memory, and TEE addresses correspond second physical cache lines. The first cache lines and the second cache lines are not shared, but separated. In this case, the attacker cannot access and fill cache lines corresponding to the TEE logical addresses. But not to limited, various schemes can be introduced about configuration of the cache memory or assignment the logical address for zero contention.

**Timing Attack Countermeasure**

In TruSpy Attack, this timing attack can be performed by sensing state change of a target cache line. Especially, the attacker tries to detect the cache line change by measuring the transition time to load cache data from the cache memory to the register. Then, we must have no difference between TEE load time and REE load time for prevent this timing side-channel attack.

**REE Data Path Delay Scheme**

Generally, TrustZone technology determines REE resource access or TEE resource access according to NS-bit. REE has longer data path latency than that of TEE because TEE performs at least one operation in a cryptography function. Then the attacker performs a timing side-channel attack by using this delay time between REE and TEE. Therefore we suggest that data path latencies of REE and TEE should be set to equal or similar to each other. In figure 7, data path of REE is configured to include a countermeasure circuit for timing attack CTA. For example, CTA may be a delay circuit (eg. flip-flop configuration) to have almost no difference in data path latency between TEE and REE.

We supposed that the shared memory of TrustZone is DRAM. Firstly, in a read operation, read data from DRAM may be one of secure data and non-secure data. The secure data is transferred to TEE via TEE data path and non-secure data is transferred to REE via REE data path. The read data from DRAM includes NS-bit, which indicates TEE data path or REE data path. MUS selects whether one of REE data path and TEE data path is connected to CPU according to the NS-bit.

Referring to Figure 7, the shown REE data path includes CTA. Accordingly data latency of REE data path is equal or similar to that of TEE data path until the read data arrive at CPU. For example, decryption time by AES decrypt is equal or similar to delay time by CTA.

Secondly, in a write operation, NS-bit indicates REE data path or TEE data path. Similarly, data latency of REE data path is equal or similar to that of TEE data path until the write data arrive at DRAM, referring to figure 8. For example, encryption time by AES encrypt is equal or similar to delay time by CTA.

**TEE Parallel Data Path Scheme**

Also, the intended data path delay in REE may have a bad effect. To solve this problem, we suggest parallel data paths in TEE. For example, due to implemented parallel crypto circuits, we can screen time difference between REE data path and TEE data path.

In read operation, non-secure data from DRAM is transferred to CPU of REE via REE data path, referring to figure 9. Also, secure data from DRAM is transferred to CPU of TEE via TEE data path having parallel AES Decryptors. AES Decryptors perform parallel decryption operation on the read data from DRAM. MUX selects one of non-secure data of REE data path and secure data of TEE data path according to NS-bit. Data of TEE data path are output from AES Decryptors. Accordingly data latency of REE data path can be same or similar to that of TEE data path until the read data arrives at CPU.

In write operation, non-secure data from CPU is transferred to DRAM of REE via REE data path, referring to figure 10. Also secure data from CPU is transferred to TEE via TEE data path h having parallel AES Encrypts. AES Encryptors perform parallel encryption operation on the write data from CPU. MUX selects one of non-secure data of REE data path and secure data of TEE data path according to NS-bit. Write data of TEE data path are outputted from AES encrypts. Accordingly data latency of REE data path can be equal or similar to that of TEE data path until the write data arrive at CPU.

**TEE Integrity Enhancement using Clark-Wilson Model**

Recently, Kaiqiang Li, Hao Feng, Yahui Li, and Zhiwei Zhang established the model of information flow control, and designed information flow control policies of MMR and Guard, which make all the communication data pass security audit and retransmission controlled by information flow control mechanism and credible components to meet the multilevel security requirements. The model and method effectively guarantee the confidentiality of the information flow in MILS using Bell-LaPadula (BLP) model [11].

The Clark-Wilson model was described by David D. Clark and David R. Wilson [10]. This Clark-Wilson model introduces a way to formalize the notion of information integrity, especially as compared to the requirements for multi-level security (MLS) systems.

Clark and Wilson argue that the existing integrity models such as Biba model were better suited to enforcing data integrity rather than information confidentiality. The Biba model is more clearly useful in, for example, banking classification systems to prevent the untrusted modification of information and the tainting of information at higher classification levels, respectively. Instead, Clark–Wilson model is more clearly applicable to business and industry processes in which the integrity of the information content is paramount at any level of classification.

The heart of the Clark-Wilson model is a relationship between a user and a set of programs (i.e., TPs) that operate on a set of data items (e.g., UDIs and CDIs). The components of such a relation, taken together, are referred to as a Clark–Wilson triple, referring to figure 11. Also the Clark-Wilson model must ensure that different entities are responsible for manipulating the relationships between principals, transactions, and data items. For example, a user capable of certifying or creating a relation should not be able to execute the programs specified in that relation. The Clark-Wilson model consists of two sets of rules: Certification Rules (C) and Enforcement Rules (E). The nine rules ensure the external and internal integrity of the data items [12].

The Clark-Wilson security model is based on preserving information integrity against the malicious attempt of tampering data. The security model maintains that only authorized users should make and be allowed to change the data, unauthorized users should not be able to make any changes, and the system should maintain internal and external data consistency [13].

**Proposed Security Access Control Model**

We assumed that the mobile device having TrustZone is a kind of a multi-level security (MLS) system. We proposed security information flow control model to enhance the integrity of the information content between REE and TEE based on the Clark-Wilson model, and built the information flow control mechanism using TrustZone driver and Authentication Tokenization Program (ATP), referring to figure 12. Generally an integrity level of a component (ex. APP) in REE is lower than that of a component (ex. TA) in TEE. Naive APP on REE cannot access a secure memory.

The proposed ATP may generate a token by communication with the TA in TEE, and transfer the token to APP. Then APP can access the secure memory using the token.

For example, if APP wants to write secure data in the secure memory based on a request, APP transforms data to be written using the token in the secure memory. For example, the transformed data may be obtained by performing a first XOR operation on data and the token, referring to figure 13. Then APP transfers the transformed data and the token to ATP.

The transformed data correspond to UDI (unconstrained data item) of Clark-Wilson Model. ATP verifies the transformed data using the token. If the verification is passed, ATP performs a second XOR operation on the transformed data and the token to obtain the data to be written in the secure memory. The data correspond to CDI (constrained data item) of Clark-Wilson Model. Then the data from ATP is transferred to the secure memory via TZ driver of Mobile OS and Trusted OS.

Also, the token is generated by TA of TEE and is stored in the secure memory via TA. ATP compares the token transferred from APP with the token stored in secure memory in the above verification operation, referring to figure 14. Also, the token may include time expiration information. ATP may perform the verification operation as to whether there is the token or whether the token is valid based on the time expiration information.

Also, ATP verifies whether the token received from APP on REE is available by comparing the token from REE with the stored token in TEE. Also, ATP can manage the token via the trusted application (TA) on TEE.

**Tokenization**

If APP on REE wants to access the secure memory, APP needs to be authenticated by ATP, firstly. ATP executes an authentication process using password, biometric information, or so on. When authenticated, ATP issues the token via TEE to enhance security of information flow from REE to TEE.

**Reading from the Secure Memory using the Token**

If APP on REE wants to read data from the secure memory, APP sends a read request with the token to ATP, referring to figure 15. ATP verifies the token of the read request by comparing the token with the stored token. The above verification is passed; ATP sends the read request via TZ to the secure OS and reads data in the secure memory in response to the read request. Then ATP performs XOR operation on the read data and the token, and sends the XOR operated data to APP.

**Writing to the Secure Memory using the Token**

If APP on REE wants to write data from the secure memory, APP sends a write request with the token and original data to ATP, referring to 15. ATP verifies the token of the write request by comparing the token with the stored token. The above verification is passed; ATP performs XOR operation on the original data and the token, sends the write request and the XOR operated data via TZ to the secure OS and writes the XOR operated data in the secure memory in response to the write request.

**Reading/Writing on TEE**

The security level of TA on TEE is equal to that of the secure memory. Then, TA has a privilege for accessing the secure memory without the token.

Proposed data flow is shown in Fig. 16. Clark-Wilson model allows high level security information to low security entity, but not allow the opposite direction to appear. Assuming that the subject S1 (ex. APP) is lower than the object 0 (ex. Secure Memory), but the subject S1 wants to write data to the object O.

At this time, the subject S1 needs to request the permission token from ATP. After the subject S1 acquires the permission token, then the subject S1 generates unconstrained data item 1 (UDI1) by a first XOR operation on data and the token. ATP may process information verification procedure (IVP). If verification is passed, APT transforms UDI1 into constrained data item 1(CDI1) by a second XOR operation on UDI1 and the token stored in the secure memory. Then, CDI1 may be written to the object O. Also, we assume that the subject S2 (ex. TA) is equal to the object 0 (ex. Secure Memory). Then the subject S2 writes CDI2 to the object O without the token.

This token-based access to TEE resource removes or mitigates OPENSSL's victim on TEE by attackers. Then, OPENSSL victim execution is reduced or failed. Accordingly, we expect to prevent TruSpy attack.

**Conclusions**

In this paper we introduced countermeasures against side-channel attacks in the shared memory of TrustZone. We proposed zero-contention cache memory or policy between REE and TEE to prevent TruSpy attacks in TrustZone. And we suggested that delay time of data path of REE is equal or similar to that of data path of TEE to prevent timing side-channel attacks. Also, we proposed security information flow control based on the Clark-Wilson model, and built the information flow control mechanism using Authentication Tokenization Program (ATP). Accordingly we can expect the improved integrity of the information content between REE and TEE on mobile devices.

**Acknowledgements**

I appreciate Ph.D. O-Kue Noh's addressing in regard with PLOS One site.

**Figure legends**

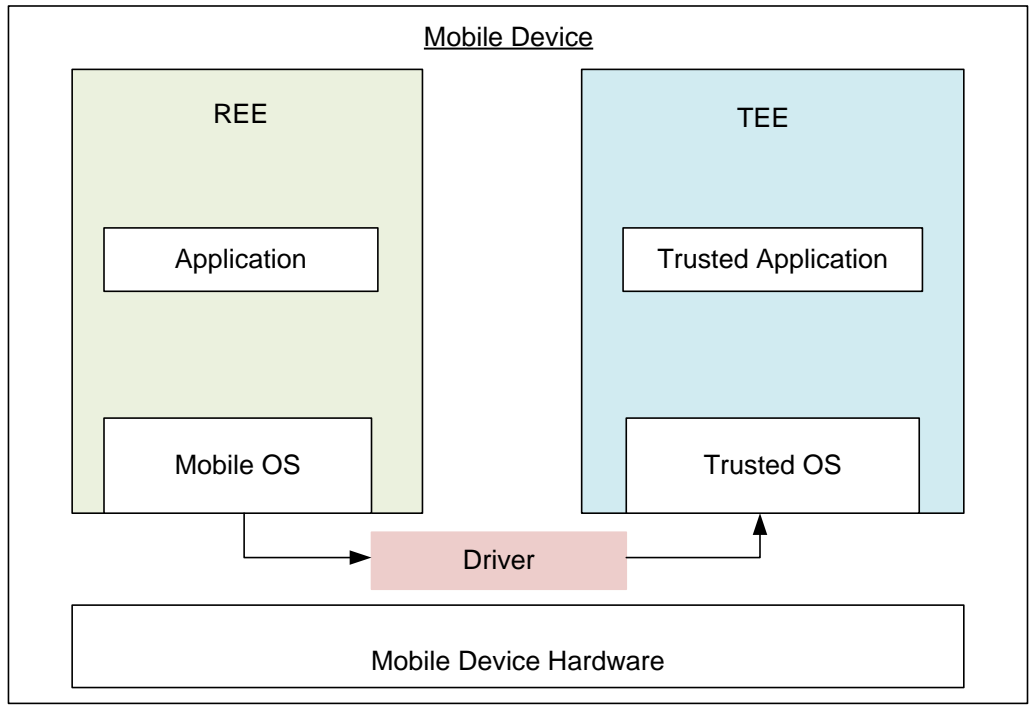

Figure 1. Mobile Device Hardware

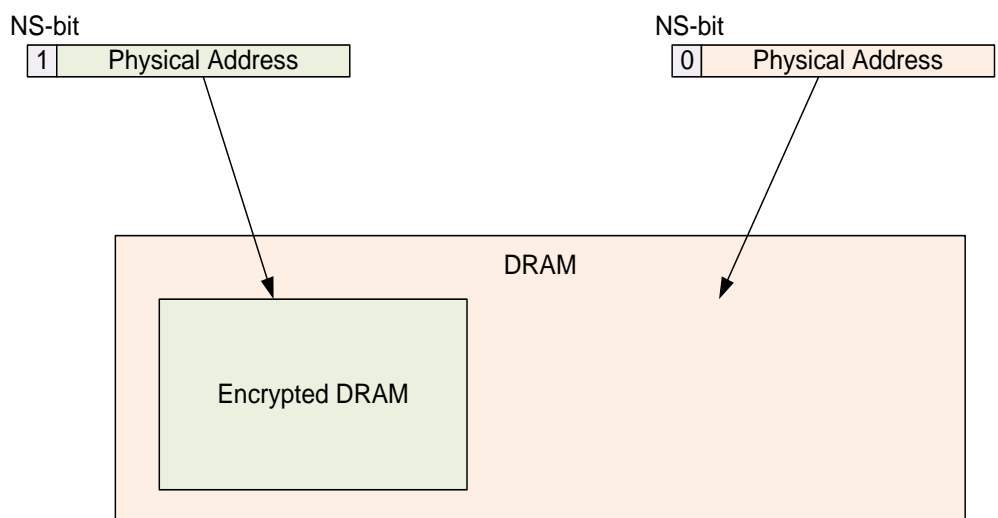

Figure 2. ARM shared DRAM

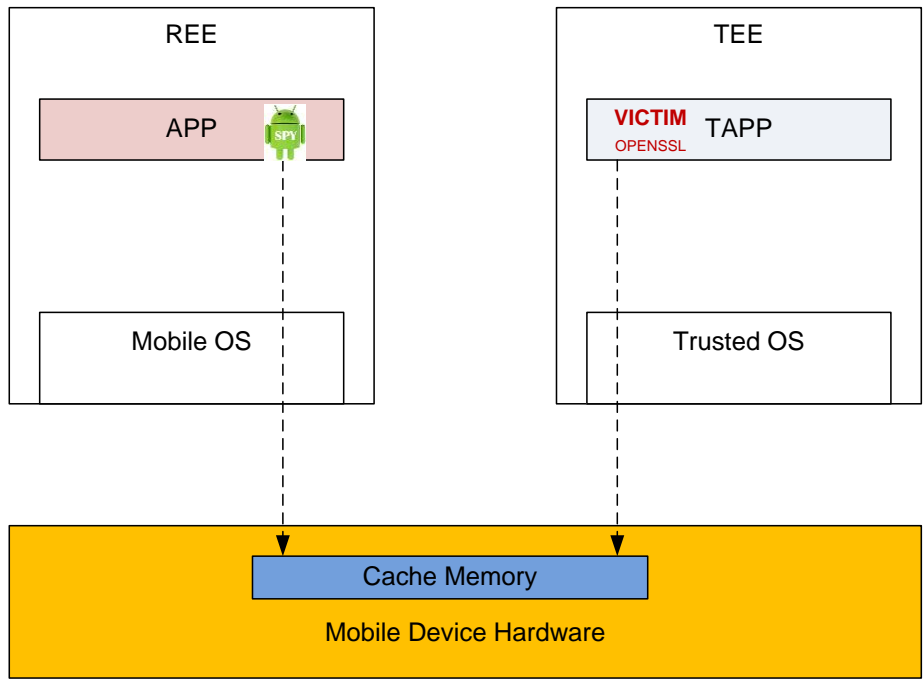

Figure 3. TruSpy Thread Model

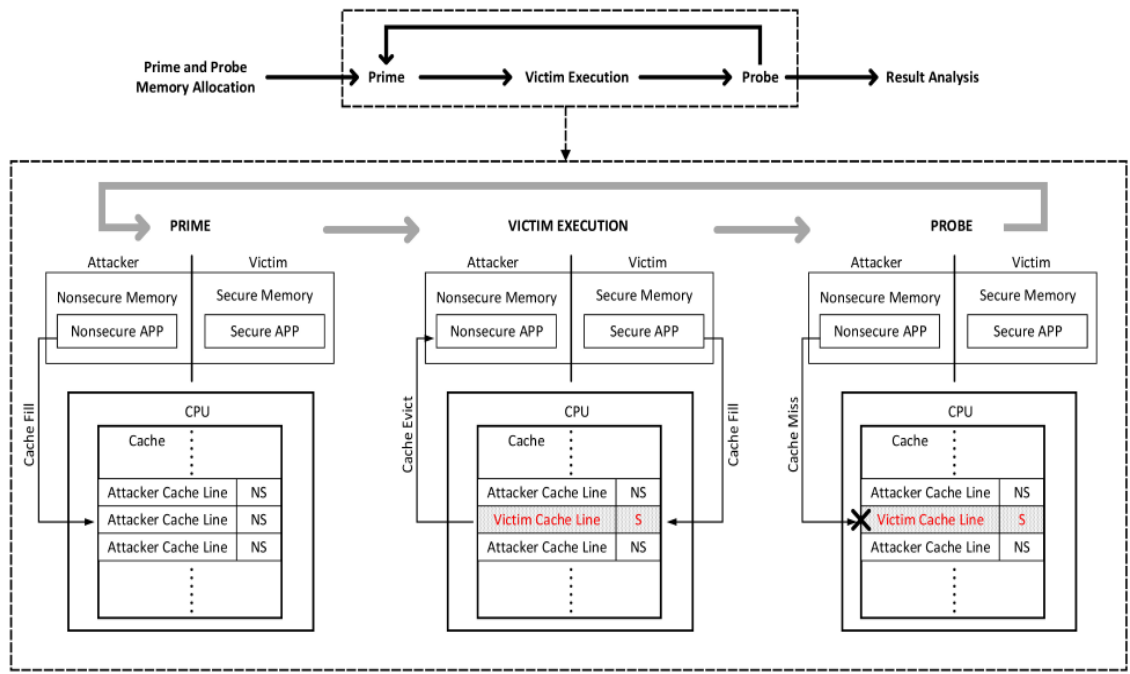

Figure 4. TruSpy Attack Flow

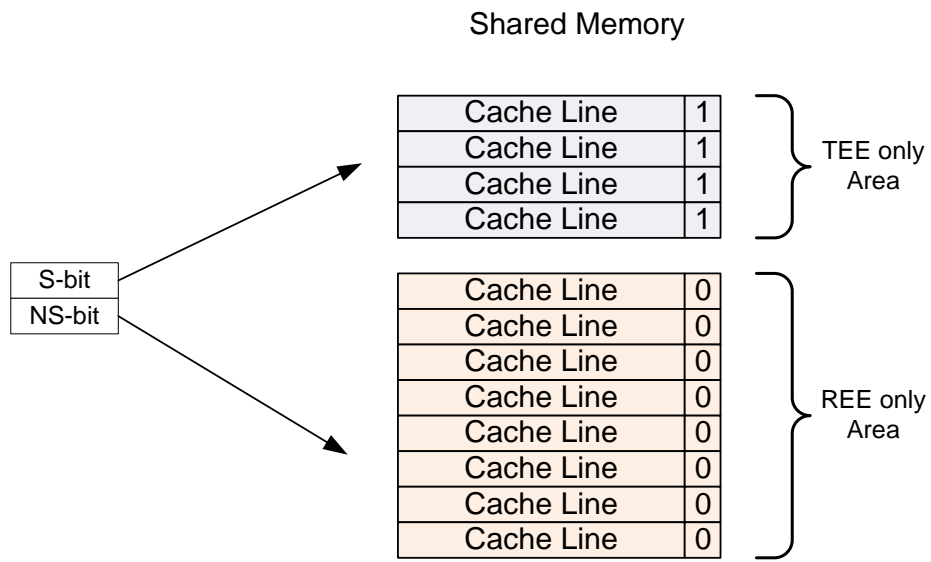

Figure 5. Hardwaredly Separate Caches

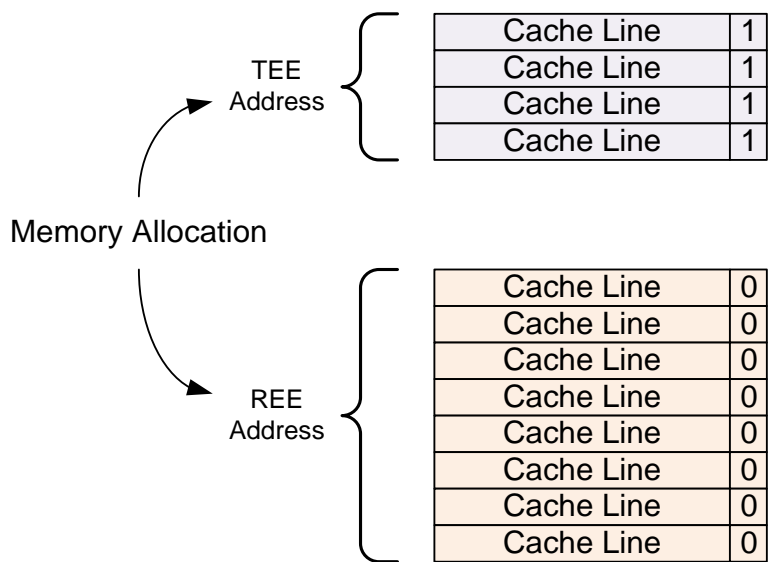

Figure 6. Separate Cache Policy

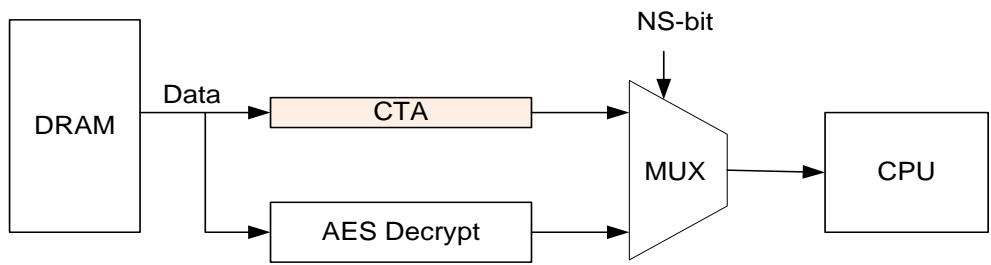

Figure 7. Read Path with CTA

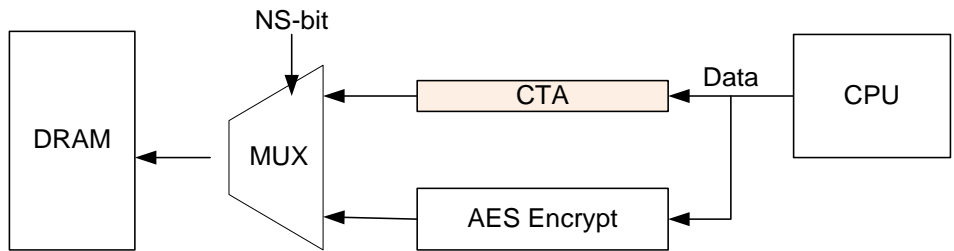

Figure 8. Write Path with CTA

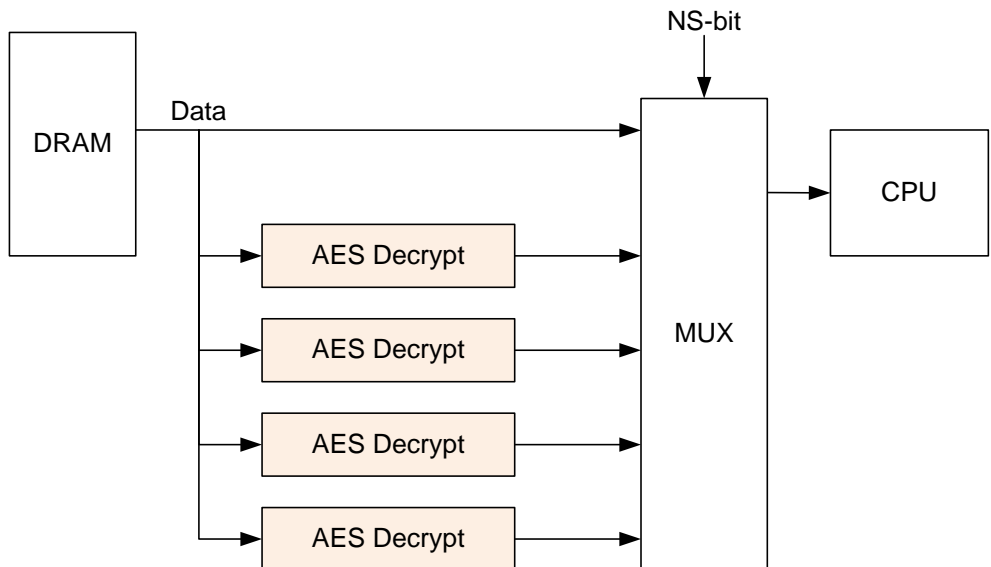

Figure 9. Parallel Decryption in Memory Read

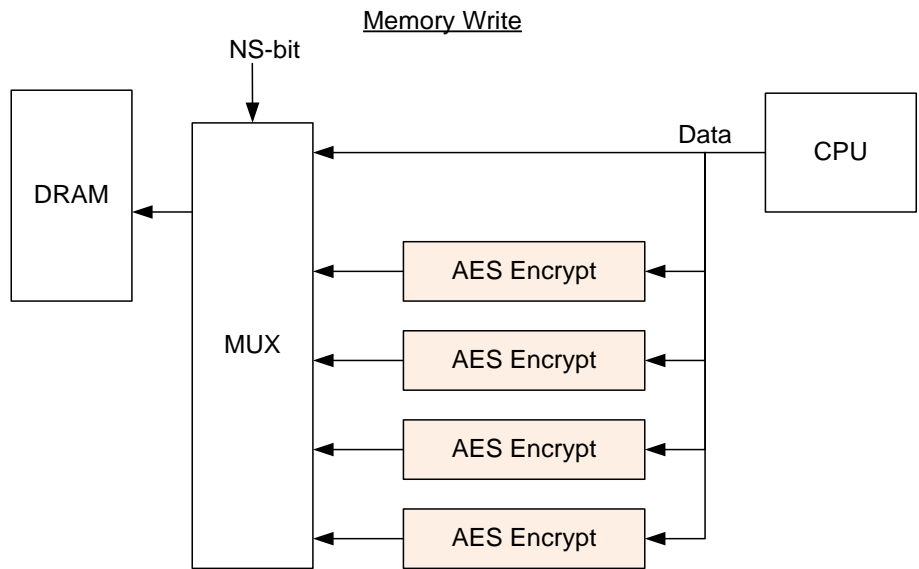

Figure 10. Parallel Encryption in Memory Write

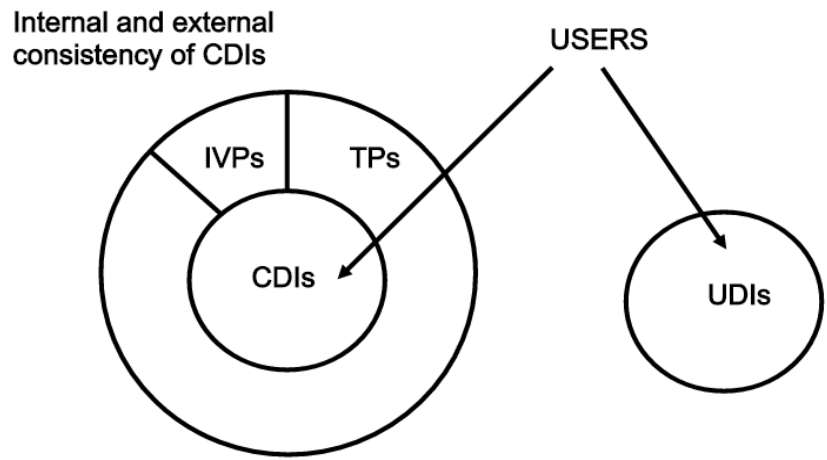

Figure 11. Clark Wilson Model

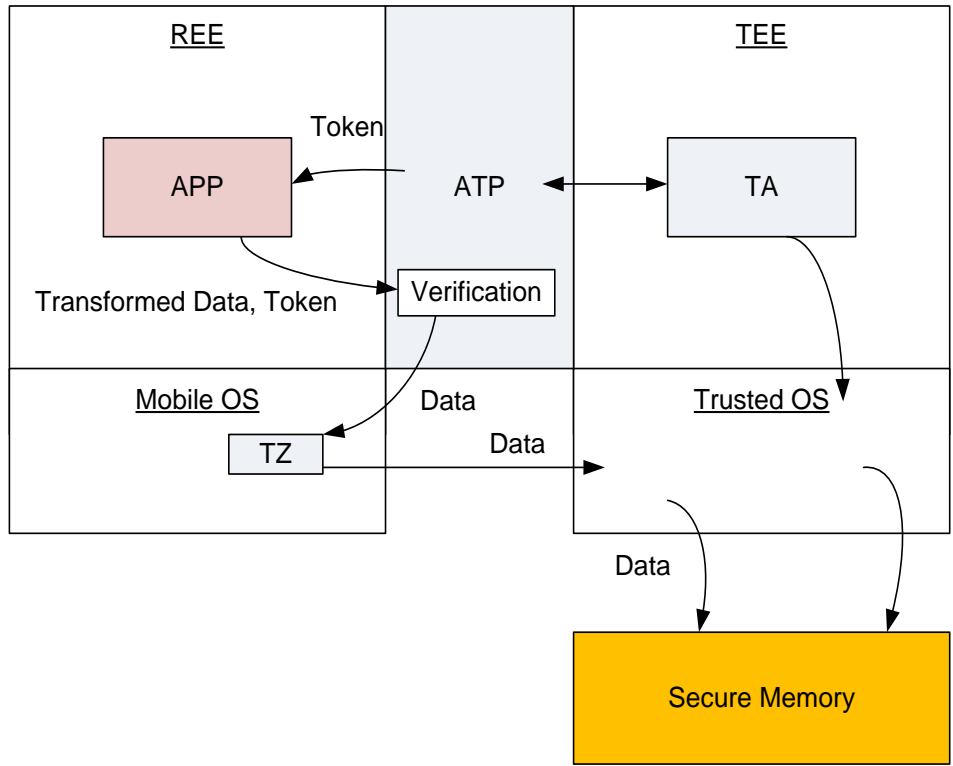

Figure 12. Proposed Access Control in Mobile Device

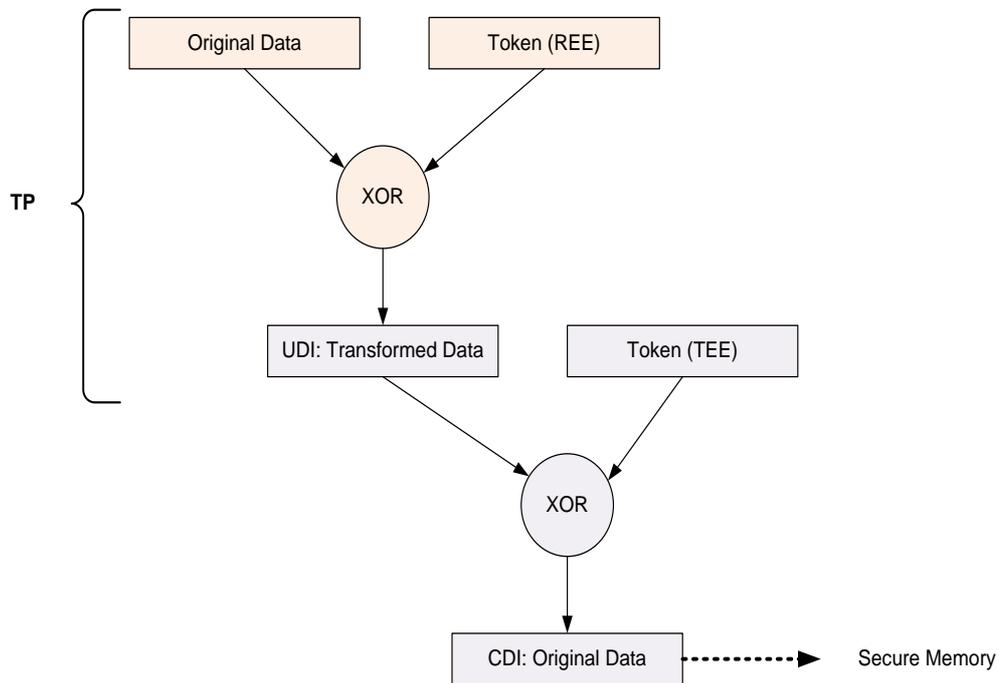

Figure 13. **Proposed Transformation Procedure (TP) based on Clark-Wilson Model**

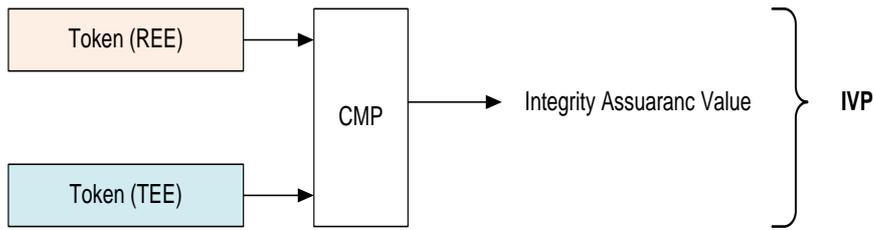

Figure 14. **Proposed Integrity Verification Procedure (IVP) based on Clark-Wislson Model**

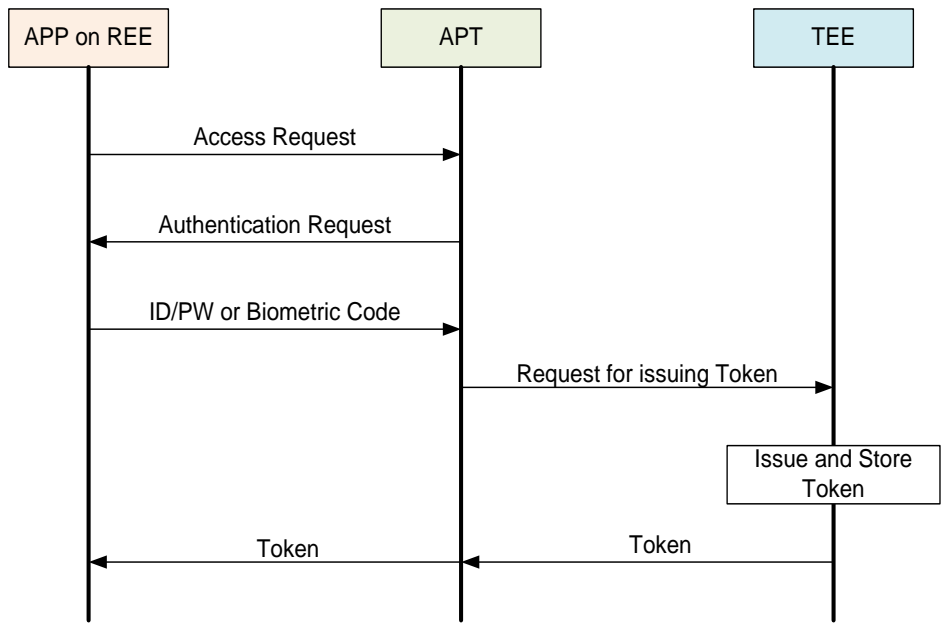

Figure 15. **Proposed Token Issuing Mechanism**

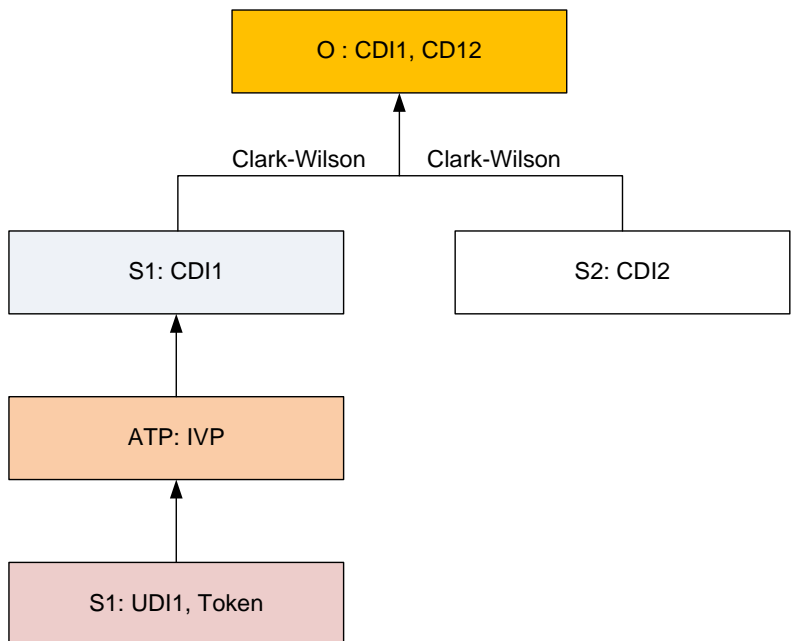

Figure 16. **Proposed Access Control Strategy on Mobile Device**